# TITLE

Great *SCO₂T*! Rapid tool for carbon sequestration science, engineering, and economics

## AUTHORS

Richard S. Middleton[1], Jeffrey M. Bielicki[2], Bailian Chen[1], Andres F. Clarens[3], Robert P. Currier[1], Kevin M. Ellett[4], Dylan R. Harp[1], Brendan A. Hoover[1], Ryan M. Kammer[4], Dane N. McFarlane[5], Jonathan D. Ogland-Hand[6], Rajesh J. Pawar[1], Philip H. Stauffer[1], Hari S. Viswanathan[1], Sean P.Yaw[7]

[1]Earth and Environmental Sciences, Los Alamos National Laboratory
[2]Civil, Environmental & Geodetic Engineering, The Ohio State University
[3]Engineering Systems and Environment, University of Virginia
[4]Indiana Geological & Water Survey, Indiana University
[5]The Great Plains Institute
[6]Department of Earth Sciences, ETH-Zurich
[7]Gianforte School of Computing, Montana State University

## ABSTRACT

$CO_2$ capture and storage (CCS) technology is likely to be widely deployed in coming decades in response to major climate and economics drivers: CCS is part of every clean energy pathway that limits global warming to 2°C or less and receives significant $CO_2$ tax credits in the United States. These drivers are likely to stimulate capture, transport, and storage of hundreds of millions or billions of tonnes of $CO_2$ annually. A key part of the CCS puzzle will be identifying and characterizing suitable storage sites for vast amounts of $CO_2$. We introduce a new software tool called *SCO₂T* (Sequestration of $CO_2$ Tool, pronounced "Scott") to rapidly characterizing saline storage reservoirs. The tool is designed to rapidly screen hundreds of thousands of reservoirs, perform sensitivity and uncertainty analyses, and link sequestration engineering (injection rates, reservoir capacities, plume dimensions) to sequestration economics (costs constructed from around 70 separate economic inputs). We describe the novel science developments supporting *SCO₂T* including a new approach to estimating $CO_2$ injection rates and $CO_2$ plume dimensions as well as key advances linking sequestration engineering with economics. Next, we perform a sensitivity and uncertainty analysis of geology combinations—including formation depth, thickness, permeability, porosity, and temperature—to understand the impact on carbon sequestration. Through the sensitivity analysis we show that increasing depth and permeability both can lead to increased $CO_2$ injection rates, increased storage potential, and reduced costs, while increasing porosity reduces costs without impacting the injection rate ($CO_2$ is injected at a constant pressure in all cases) by increasing the reservoir capacity. Through uncertainty analysis—where formation thickness, permeability, and porosity are randomly sampled—we show that that final sequestration costs are normally distributed with upper bound costs around 50% higher than the lower bound costs. While site selection decisions will ultimately require detailed site characterization and permitting, *SCO₂T* provides an inexpensive screening tool that can help prioritize projects based on the complex interplay of reservoir, infrastructure (e.g., proximity to pipelines), and other (e.g., land use, legal) constraints on the suitability of certain regions for CCS.

## KEYWORDS

# INTRODUCTION

$CO_2$ capture and storage (CCS) technology is likely to be widely deployed in coming decades due to major climate drivers (CCS is part of every major climate policy that limits global warming to 2°C) and economic drivers (significant $CO_2$ tax credits in the United States). These drivers are expected to stimulate capture, transport, and storage of hundreds of millions or billions of tonnes of $CO_2$ annually (100s $MtCO_2$/yr to 1+ $GtCO_2$/yr). Sequestering large amounts of $CO_2$ requires identifying hundreds or thousands of potential storage sites and understanding how geological characteristics (such as formation depth, thickness, permeability, porosity, and temperature) and logistical parameters (such as well patterns/spacing and brine treatment/disposal) will impact sequestration potential and associated costs [1]. Despite this, the science and approach to identifying adequate sequestration sites on this scale—billions of tonnes of $CO_2$ annually—do not exist.

To meet this challenge, we introduce a fast-running tool called *SCO₂T* (Sequestration of $CO_2$ Tool, pronounced "Scott"). The tool that can rapidly screen hundreds of thousands of reservoirs, perform sensitivity and uncertainty analysis, and estimate geologic $CO_2$ storage costs. *SCO₂T* is the first tool that comprehensively and directly links key outputs from sequestration simulations with detailed sequestration economics. Consequently, the tool can explore complex and potentially counterintuitive relationships of reservoir characteristics, such as the impact of increasing depth, thickness, permeability, porosity, and temperature on sequestration engineering and costs.

In addition to providing an approach to understand how sequestration costs can vary across storage site parameters including geology and logistics, *SCO₂T* estimates site-wide reservoir capacities by calculating the amount of $CO_2$ that can be realistically or actually injected and stored in a given 3D geological block. This takes into account injection rates, the number of wells and well patterns, and plume dimensions. These are used to calculate how much $CO_2$ can actually be stored. Contrariwise, other site-wide approaches. such as the FE/NETL $CO_2$ Saline Storage Cost Model [2], typically use a storage efficiency [3] and other calculations to estimate storage capacity and don't focus on the impact of injection rates and plume characteristics. *SCO₂T* is underlain by full-physics sequestration simulations, including pressure- and temperature-dependent multiphase flow (FEHM [4-6]), which allows it to rapidly calculate outputs without sacrificing detailed accuracy.

This paper proceeds as follows. First, in *Background*, we first present a literature review of research that our work has grown out from. Second, in *Approach*, we introduce the *SCO₂T* framework including background information and work on *SCO₂T* inputs and outputs, the modified ROMster approach [7], a detailed overview of the economic inputs and calculations, and an overview of the *SCO₂T* framework. This *Approach* section includes highlighting some of the novel scientific advances that were required to build *SCO₂T*. as well as validation. Third, in *Results and Discussion*, we present a sensitivity analysis and uncertainty analysis of multiple carbon sequestration scenarios. Fourth, in *Conclusions and Future Research*, we highlight the major findings from our study and outline future research directions. All data used in this study (model runs and data outputs for the figures) are included in a supporting Microsoft Excel file including all original figures.

# BACKGROUND

In recent years, multiple approaches, models, and tools have been developed for performance assessment of geologic $CO_2$ sequestration (GCS). Here, we provide some of the most relevant works to put the *SCO₂T* tool and its supporting science in context. Stauffer et al. [8, 9] introduced a system model called *CO$_2$-PENS* (Predicting Engineered Natural Systems) for sequestering CO$_2$ in geologic reservoirs based on the GoldSim [10, 11] platform. GoldSim is a system-modeling package which is designed for stochastic modelling applications of engineered geologic systems (such as GCS), particularly those with large uncertainties. *CO$_2$-PENS* was designed to conduct probabilistic simulations of CO$_2$ capture, transportation, and injection in different geologic formations. It can be used to explore relationships between uncertain variables and can help to distinguish the likely performance of potential sequestration sites. In addition, *CO$_2$-PENS* has the capability to link an injection module with a simple economic module and was modified [12] for use in the *SimCCS* [13-18] CCS decision support tool. Zhang et al. [19] developed a system-level model for GCS including CO$_2$ capture, compression, CO$_2$ transportation, and injection which was also based on GoldSim. Oldenburg et al. [20] developed a certification framework (CF) for certifying the safety and effectiveness of GCS sites; CF relates effective CO$_2$ trapping to leakage risk. Through the generality and flexibility, the CF approach can help with the assessment of CO$_2$/brine leakage risk as part of the certification process for permitting of GCS sites. Metcalfe et al. [21] developed a generic system model using Quintessa's QPAC software [22], and the model was then adapted and applied to the demonstration of CO$_2$ storage at Krechba, near In Salah in central Algeria. Most recently, the US DOE-funded National Risk Assessment Partnership (NRAP) developed an integrated assessment model (NRAP-IAM-CS [23]) that can be used to simulate CO$_2$ injection, migration, and associated impacts (e.g., potential geochemical impacts to groundwater) at GCS sites. NRAP-IAM-CS incorporates a system-modeling-based approach which accounts for the full subsurface system from the storage reservoir to groundwater aquifers and the atmosphere. NRAP has recently released on open source IAM called NRAP-Open-IAM which provides similar functionality to NRAP-IAM-CS, but draws on a larger set of model analysis tools, is cross-platform, and can execute simulations concurrently on parallel computational resources [24]. Except for *CO$_2$-PENS*, none of the above approaches or models can directly link key outputs (CO$_2$ injection rates, plume dimensions, etc.) from sequestration simulations with sequestration costs or economics.

## APPROACH

*SCO$_2$T* uses a set of reduced-order models (ROMs) to calculate two key sequestration engineering outputs for any given formation: CO$_2$ injection rate and CO$_2$ plume area. These engineering outputs are then used with other inputs to calculate the annualized cost of geologic CO$_2$ storage. Because *SCO$_2$T* is a rapid sequestration science and screening tool, it is not designed to replace detailed sequestration modeling of individual sites, where fine-scale reservoir heterogeneity and fluid characteristics have substantial impacts on the storage estimates and injection well patterns and designs. Consequently, *SCO$_2$T* makes several high-level assumptions including reservoir homogeneity and does not consider issues such as leakage [25-27] or reservoir fluid composition [28].

### *Inputs and outputs*

*SCO$_2$T* requires five pieces of geologic information to characterize a reservoir for any one realization: formation depth, thickness, permeability, porosity, and geothermal gradient or temperature. These are described in more depth in the supporting information (SI). These five pieces of geologic information are used as inputs to the reduced-order models described in the next section. *SCO$_2$T* needs further information to calculate site-wide sequestration engineering and economics including reservoir 2D area (km$^2$), maximum injection pressure as a fraction of lithostatic pressure (currently cannot be changed from a default value of 0.8) based on a grain density of 2650 kg/m³ and porosity of 0.15 (fraction) for the rocks overlying the storage, maximum well injection rate (currently cannot be changed and is assumed to be 1 $MtCO_2$/yr per well), and injection period (in years; current version only allows a 30-year period). *$SCO_2T$* also requires additional inputs including economics (a capital charge factor), the cost of treating/disposing a cubic meter of brine ($/m³; the value can be zero), the number/proportion of brine extraction wells relative to injection wells needed to be drilled for each injection well (can be zero), whether or not a pump is needed for each $CO_2$ injection well (if the $CO_2$ arrives at the site at sufficient pipeline pressure, a pump may not be required), and well construction design including whether fractional wells can be drilled (useful for sensitivity analysis where you could place 1.1 wells in a small area), square (aligned) vs. hexagonal (staggered) well spacing (hexagonal is the default), and whether $CO_2$ plumes can overlap.

*$SCO_2T$* uses the five geologic inputs in combination with the calculated maximum injection pressure to calculate the maximum injection rate ($MtCO_2$/yr), plume radius (km), and plume volume (million cubic meters or Mm³) for an individual injection well. Here, plume volume is used interchangeably with "injected $CO_2$ volume" and does not take into account rock pore space (i.e., plume volume is the mass of $CO_2$ divided by the $CO_2$ density). For cases where the maximum injection rate exceeds 1 $MtCO_2$/yr, *$SCO_2T$* recalculates the plume radius and volume using the maximum injection rate of 1 $MtCO_2$/yr. Even though many combinations of geologic parameters could conceivably exceed 1 $MtCO_2$/yr, typical well construction, including standard casing and drilling technology, do not typically permit flow rates in excess of 1 $MtCO_2$/yr. These individual well outputs are then translated into site-wide outputs. First, the site-wide area is divided by the individual plume areas to calculate how many wells can be placed at the site. This considers whether wells are hexagonally or squarely spaced and if plumes can overlap. Second, the injection rate and number of wells are combined to calculate a site-wide reservoir capacity. Third, the number of wells and site-wide values are translated into a detailed series of individual economic estimates that are integrated to generate a final sequestration cost ($/$tCO_2$).

### *Reduced order models (ROMs)*

*$SCO_2T$* uses a set of reduced-order models (ROMs) to calculate two key sequestration engineering outputs for any given formation: (1) $CO_2$ injection rate per well and (2) $CO_2$ plume area (results are often reported as a radius in this paper). The plume area can also be calculated using the Nordbotten [29] analytical solutions rather than the ROMs; this is largely available for reference purposes. ROMs are a widely used and powerful approach to reducing the complexity of predictive physics-based numerical simulations in GCS [30-32]. They allow fast computations of entire system performance even for periods of hundreds to thousands of years. Traditional approaches generate a single ROM for each simulated responses (e.g., $CO_2$ injection rate) based on a set of training simulations [33-35]. Chen et al. [7] demonstrated that a single ROM can display excellent overall predictive statistics, but have predictions that dramatically and unacceptably deviate from simulator responses. To avoid the potential pitfall of traditional ROMs development, Chen et al. [7] developed a ROMster approach and showed how this approach can reduce the average absolute error from 200% to only 4% in their study.

The current version of *$SCO_2T$* calculates the injected $CO_2$ volume by dividing the injected mass by the density of the $CO_2$ in the reservoir (i.e., plume volume is the same as injected $CO_2$ volume). Injected volume is used to estimate how much brine might be displaced during $CO_2$ injection and storage and, subsequently, how much could be extracted and disposed or treated. The *$SCO_2T$* ROMs were trained using a set range for the five geologic inputs—formation depth, thickness, permeability, porosity, and geothermal gradient—based on the reasonably expected range for potential sequestration sites (Table 1). ROMs, particularly using spline fitting, have no guarantee of performing well outside their trained range and often have difficulties when they are approaching the bounds of their trained range [7]. Consequently, the ROMs were trained for a ±10% range for each geologic input beyond their "operational" range (slightly larger lower bounds for thickness and porosity). Training ranges and user ranges are listed in Table 1. Full details of how the injection rate and plume area ROMs were developed are presented in the SI.

| Parameter | Units | Training range | | User range | |
|---|---|---|---|---|---|
| | | Lower | Upper | Lower | Upper |
| Depth | m | 900 | 5500 | 1000 | 5000 |
| Thickness | m | 4.5 | 110 | 5 | 100 |
| Permeability | mD | 0.9 | 1100 | 1 | 1000 |
| Porosity | fraction | 0.03 | 0.44 | 0.05 | 0.4 |
| Geothermal gradient | °C/km | 13.5 | 49.5 | 15 | 45 |

Table 1: Geologic input parameters for the $SCO_2T$ ROMs and their training and operational ranges.

## *Economics*

Sequestration economics is perhaps the ultimate driver for identifying suitable $CO_2$ sites; issues such as environmental impact and risks are vitally important, but economic issues are arguably a showstopper for sequestration feasibility. For example, a site with expected high sequestration costs would arguably not pass the financing stage. Typically, sequestration costs are only calculated on a site-by-site basis within individual projects. There is less broad understanding of how sequestration costs vary across storage site parameters including geology (e.g., formation depth, thickness, permeability, porosity, and temperature) and logistics, such as well spacing and brine treatment/disposal costs. Well spacing is described in depth in the SI. This broad understanding is critical for site screening, which is often the first step in identifying storage potential of different geographical regions or reservoirs. Screening activity might involve examining hundreds or even many thousands of potential sites to down-select suitable candidates for deeper exploration. No tool exists for this down-selection process currently and is an engineering and science gap in $CO_2$ sequestration science. Screening also often involves sensitivity analysis, whereby a user seeks to understand the impact of one or more parameters on sequestration potential. $SCO_2T$ addresses both of these challenges.

$SCO_2T$ directly links key outputs from sequestration simulations ($CO_2$ injection rates, plume dimensions, displaced brine volumes, etc.) with comprehensive sequestration costs or economics. Although other tools have done this too, such as *$CO_2$-PENS*, $SCO_2T$ is the first approach to take realistic physics-based estimations of $CO_2$ injection and storage and link this with detailed economics. Consequently, the tool can explore complex and potentially counterintuitive relationships of reservoir characteristics, such as the impact of increasing depth, thickness, permeability, porosity, and temperatures on engineering and costs. The $SCO_2T$ approach of connecting sequestration simulations or engineering with detailed economics was first developed in Middleton et al. [12] and preliminary versions used in subsequent studies. These preliminary versions were used to parametrize datasets for the CCS infrastructure model *SimCCS* [13-17] and to conduct studies of the impact of geology on sequestration costs. For example, Middleton and Yaw [36] examined 20 potential storage reservoirs in Alberta, Canada, identifying the impact of uncertainty (formation thickness, permeability, and porosity) on storage costs including a threshold where injectivities below 0.25 MtCO$_2$/yr leads to exponentially rising costs. Further details on the economics approach are detailed in the SI.

## *SCO$_2$T* framework

*SCO$_2$T* (Figure 1) is developed in Microsoft Excel and the code (2000+ lines) written in Visual Basic for Applications (VBA). Excel was chosen for multiple key reasons: the vast majority of users already have access to Excel, no installation is required, it's useable on PC and Mac systems, Excel encapsulates the data in an easy-to-read/enter format *and* the open-source coding language within a single platform, and it has in-built capability to visualize outputs (each execution of *SCO$_2$T* produces 81 individual charts exploring outputs). *SCO$_2$T* is also fast running: even though VBA is an un-compiled language, *SCO$_2$T* typically performs 10,000 separate realizations in a second on a single processor of a typical modern PC or Mac. For comparison, the original 10,000 FEHM training simulations, used to train the *SCO$_2$T*'s ROMs, took on average five hours for a single simulation with many taking several days [7]. The fast-running capability is important for site screening. For example, *SCO$_2$T* could perform a single simulation of all 186,675 10x10 km grid cells in the NATCARB Atlas [37] in less than 30 seconds. Speed is important for sequestration site screening as well, where hundreds of thousands or millions of simulations to support sensitivity and uncertainty analysis can be performed in minutes. In addition to printing output to the *SCO$_2$T* tool itself (i.e., the main Worksheet), *SCO$_2$T* can export the data to a text file or export directly into a format to be read by *SimCCS* [17]. *SCO$_2$T* is being publicly released with this manuscript.

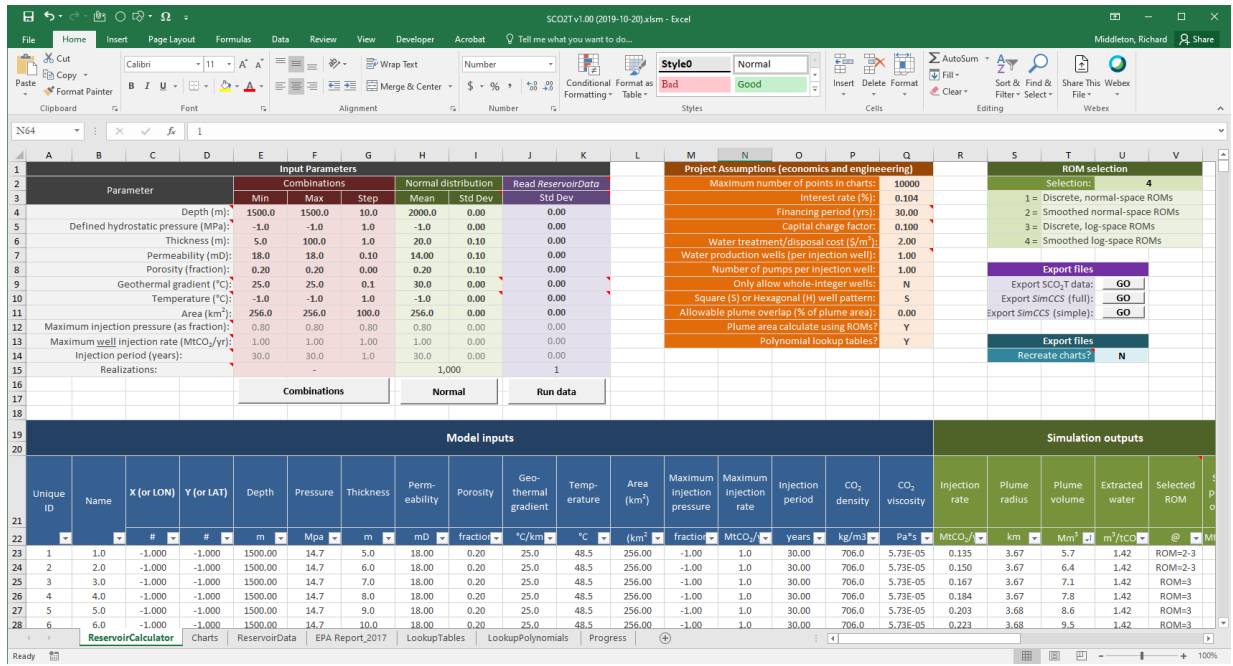

Figure 1: Screenshot of the *SCO$_2$T* interface.

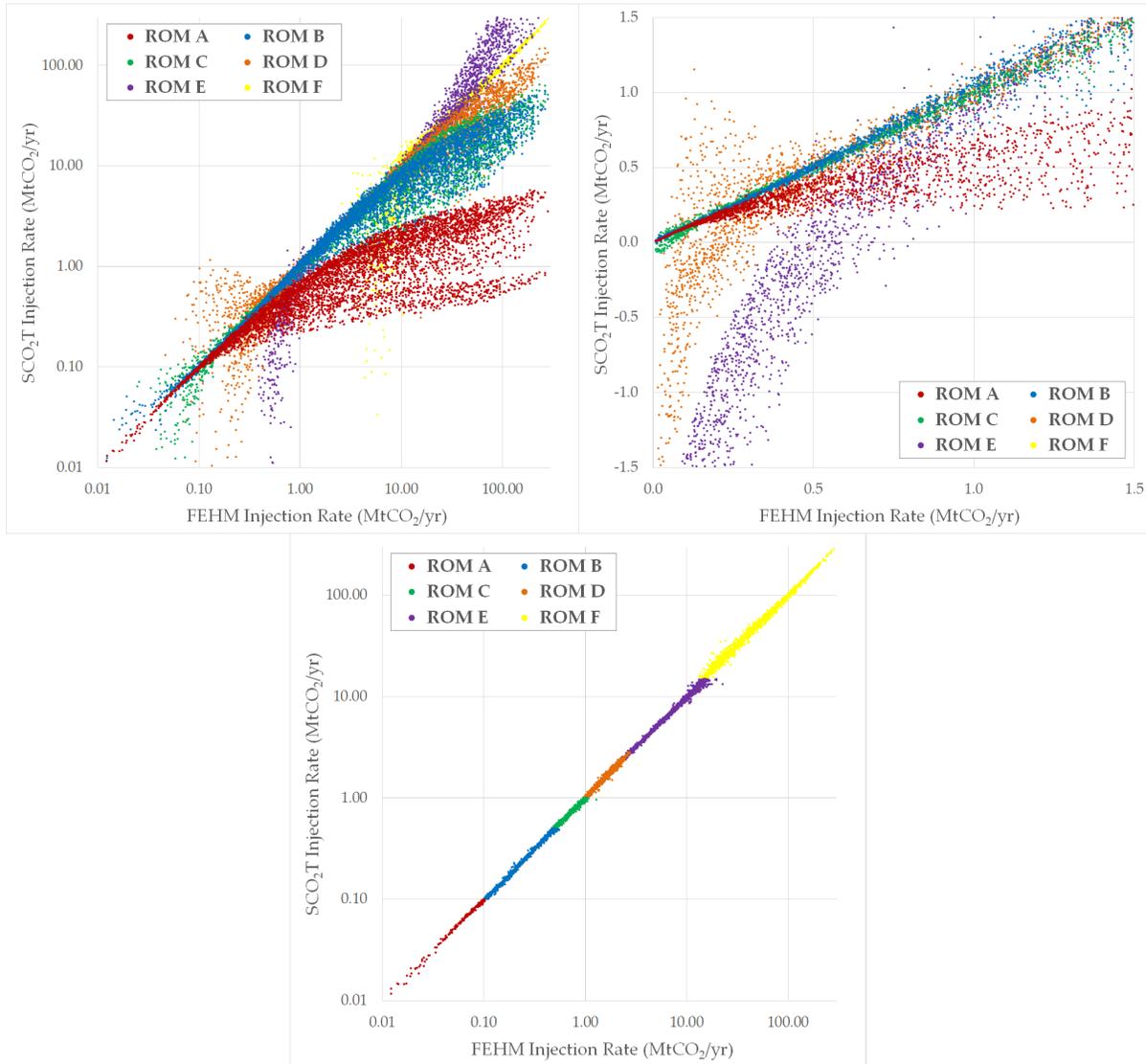

Figure 2: CO$_2$ injection validation of the ROMster approach in *SCO$_2$T* using all training data within the useable range: (left) performance of the six injection rate ROMs across the entire injection range; (middle) six-ROM performances focused on injection rates below 1.5 MtCO$_2$/yr; and (right) final "merged" ROMs using the ROMster approach. Note that left and right plot axes are in log scale while the middle plot axes are in natural scale.

## Validation

*SCO$_2$T* utilizes the novel ROMster approach to predict CO$_2$ injection rate and plume area given formation depth, thickness, permeability, porosity, and temperature. Figure 2 illustrates the success of the ROMster approach for the six CO$_2$ injection ROMs across the entire injection space defined by the 10,000 FEHM simulations. None of the six ROMs perform well over large parts of the injection space (Figure 2a/b), let alone the whole range, though each ROM can be seen to perform well in their trained range (Table SI in the SI). As ROMs move outside of their trained injection range, their prediction success rapidly deteriorates with errors reaching two or more orders of magnitude. For example, in Figure 2a it can be seen that ROM A (red dots) has very high accuracy in its trained range—it's trained on FEHM runs that led to an injection rate of 0–0.1 $MtCO_2$/yr—given by the 1:1 line between the FEHM output (x-axis) and the *$SCO_2T$*-predicted injection rate (y-axis). However, when ROM A is used with combinations of depth, thickness, permeability, porosity, and temperature that lead to an FEHM injection rate greater than 0.1 $MtCO_2$/yr, the ROM performance dramatically pulls away from the 1:1 relationship. The same is true for the other five ROMs. For example, ROM B (green dots, trained on 0.05 to 0.5 $MtCO_2$/yr injection rates) performs poorly both below 0.05 $MtCO_2$/yr (both over- and under-predicting injection rates) and above 0.5 $mCO_2$/yr (under-predicting injection rates).

These errors are particularly pronounced in the key injection range of 0.25–1 $MtCO_2$/yr. This range is key because injection rates below 0.25 $MtCO_2$/yr are likely to lead to prohibitively high injection costs and injection rates are capped at 1 $MtCO_2$/yr based on well dimensions. That is, real-world geologic carbon sequestration is likely to happen in this range. Figure 2b plots combinations of ROM performance for the FEHM-predicted range 0–1.5 $MtCO_2$/yr (x-axis) and the *$SCO_2T$*-predicted range -1.5 to 1.5 $MtCO_2$/yr. Focusing on this key injection range shows that ROMs quickly perform badly outside their range even to the extent where all ROMs apart from ROM A (trained on the range 0–0.1 $MtCO_2$/yr) can predict negative $CO_2$ injection rates! Using the ROMster approach, the six ROMs can be merged to provide an excellent prediction of $CO_2$ injection rates across the entire range (Figure 2c) with an $R^2$ of 0.9989. Even with such high predictive power, there can be minor gaps between ROMs. For example, a sensitivity analysis of increasing depth from 1000 m to 5000 m means *$SCO_2T$* will have to jump from making predictions with ROM A to ROM B. Consequently, *$SCO_2T$* support so that the user can either decide to always use one discrete ROM for the prediction or, for injectivities that fall between any two ROMs, use a weighted average of predictions (this is the suggested and default approach).

Figure 3 illustrates the validation of the $CO_2$ plume area ROMs. The blue points and linear regression equation in Figure 3 (left) indicate that the ROMs reproduce the FEHM outputs with high accuracy (note the $R^2$ value and the 1:1 relationship). The ROMs are not colored separately for each of the three plume area ROMs because, unlike for injection rate, the ROMs all overlap (e.g., a large plume area could indicate a high injection rate in thick formation or a low injection rate in a thin formation). The red dots refer to the Nordbotten [29] analytical solution for plume evolution that formed the basis for plume radius used originally in *$CO_2$-PENS [8]*. *$SCO_2T$* allows the user to calculate the plume area using either the Nordbotten analytical solution or the ROM. Even though the Nordbotten approach uses simplifying assumptions in a first principles derivation, the analytical solution performs fairly well compared with *$SCO_2T$* for unlimited injection rates. Figure 3 (right) shows the validation for the *$SCO_2T$* and Nordbotten plume area approaches for the rate-controlled simulations. Nordbotten solutions generally have a much greater scatter from the 1:1 reference and, due to simplifying assumptions, are not able to capture the complexity of interactions between the five major geologic parameters.

Currently, *$SCO_2T$* estimates the $CO_2$ injected volume by dividing the total injected mass of $CO_2$ by the density of the stored $CO_2$. Future versions will potentially create a separate $CO_2$ plume volume set of ROMs. The plume volume is used to calculate the amount of brine produced (taking into account $CO_2$ and brine density) to keep the reservoir at hydrostatic pressure. The volume of the injected $CO_2$ is also required for Nordbotten's plume area calculation along with $CO_2$ and water viscosity. $CO_2$ and water density and viscosity lookup tables are included with the *SCO₂T* tool for temperatures ranging from 0.5°C to 300°C (every 0.5°C) and pressures ranging from 0.25 to 60 MPa (every 0.25 MPa); these were specifically developed for *SCO₂T* using the NIST Chemistry WebBook [38]. Increasing temperature reduces $CO_2$ density and generally decreases the viscosity (though not always; see Figure S6), while increasing pressure increases both $CO_2$ density and viscosity. To find the density and viscosity for $CO_2$/water, typical approaches bilinearly interpolate between temperatures and pressures in the lookup table. However, $CO_2$ density and viscosity both respond nonlinearly to changes in temperature and pressure and so linear averages introduce errors (see SI). To address this, *SCO₂T* allows the option to use a polynomial weighting that almost entirely removes this inaccuracy (see Figure S7 and explanation in the SI); this is the suggested and default option.

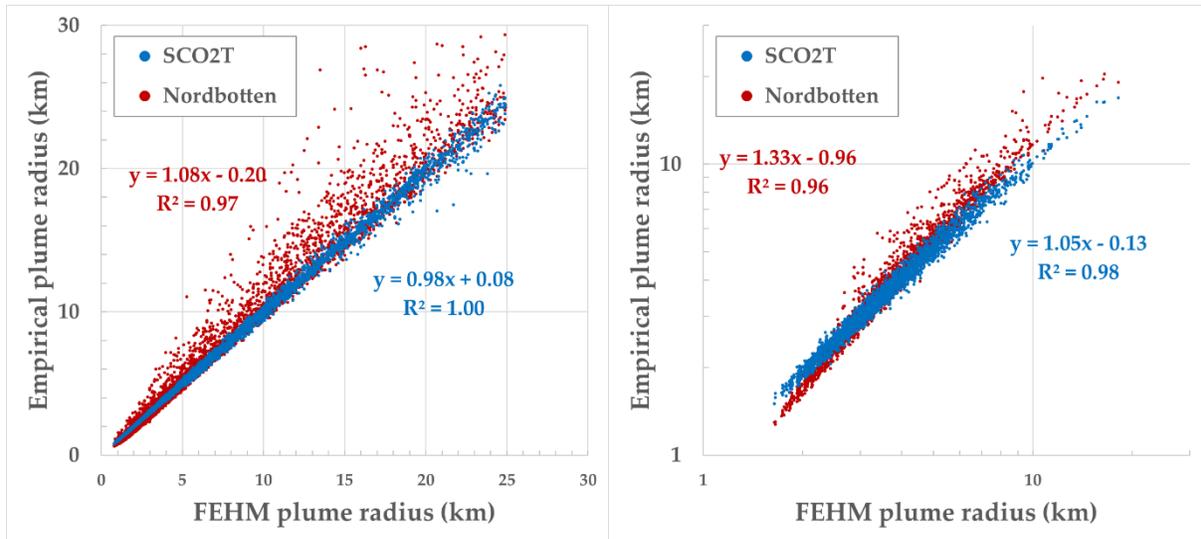

Figure 3: $CO_2$ plume area validation of the ROMster approach in *SCO₂T* for $CO_2$ plume radius: (left performance of the three plume area estimates across the entire injection range; and (right) ROM performance when the unlimited injection rate is above 1 MtCO₂/yr but the rate is limited to 1 MtCO₂/yr. Blue points are the performance of the *SCO₂T* ROMs and red points are the plume areas calculated using Nordbotten's analytical approach.

# RESULTS AND DISCUSSION

## Sensitivity analysis

The following two sections summarize sensitivity and uncertainty analyses to demonstrate the power of *SCO₂T* and to identify key $CO_2$ sequestration issues as well as the impact on economics. The five geologic inputs—depth, thickness, permeability, porosity, and temperature—each impact $CO_2$ injection rates, plume dimensions, storage effectiveness, and extracted brine in different ways, and therefore with differing and sometimes counterintuitive impacts on economics. For sensitivity analysis, four of the five variables are typically held constant while the remaining variable is varied to understand its impact. This is a local or transect sensitivity analysis rather than a full global sensitivity analysis.

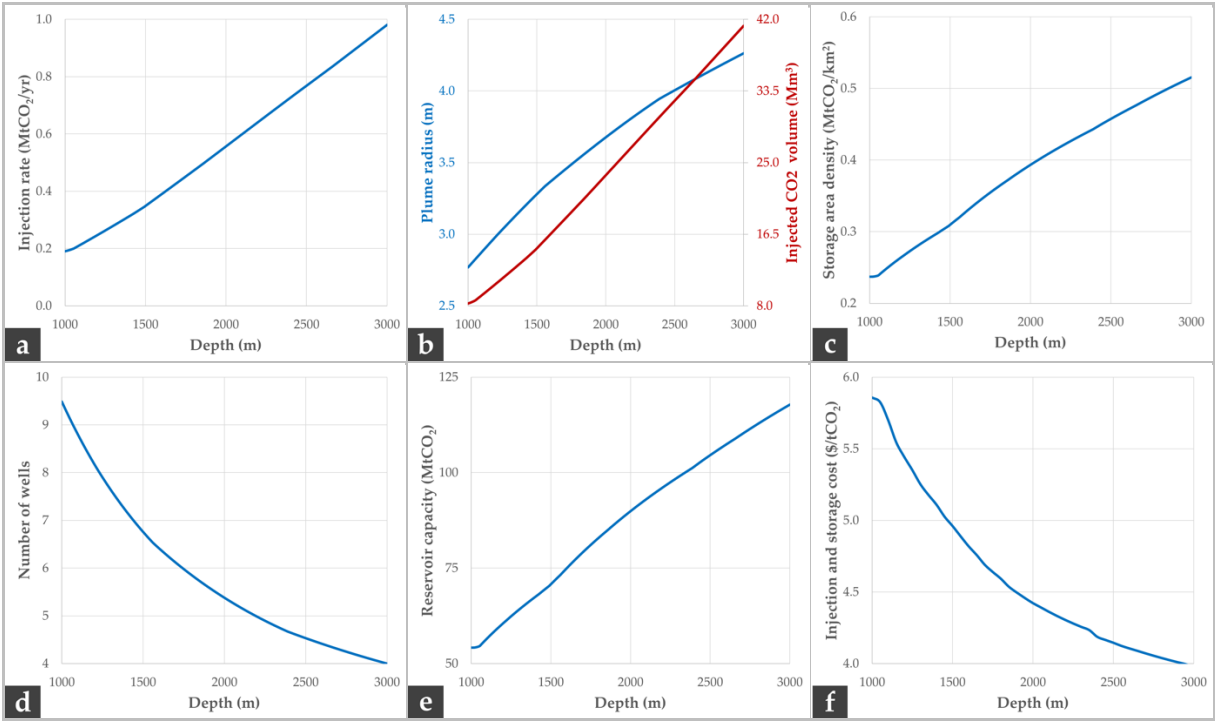

Figure 4: Impact of varying **depth** from 1000 to 3000 m. Remaining variables: Thickness = 20 m; permeability = 14 mD, porosity = 0.2, geothermal gradient = 25°C/km. By default, *SCO₂T* assumes that each reservoir is 256 km² (roughly 10 by 10 miles).

Figure 4 illustrates the impact of holding thickness, permeability, porosity, and geothermal gradient constant and varying depth from 1000 to 3000 m. Increasing formation depth increases both formation temperature and pressure. Increasing pressure increases both the $CO_2$ viscosity and density, while increasing temperature reduces $CO_2$ density and viscosity. These changes subsequently impact $CO_2$ injection rate (Figure 4a), $CO_2$ plume radius and volume (Figure 4b), $CO_2$ storage area density (Figure 4c), number of wells (Figure 4d), reservoir capacity (Figure 4e), and total injection and storage cost (Figure 4f). Storage area density is a measure of the mass of $CO_2$ that can be effectively stored in the reservoir normalized by the reservoir's area ($MtCO_2/km^2$)—as opposed to the storage capacity of an entire reservoir which is simply based on the overall pore space of the reservoir without considerations of the actual ability to utilize that space—and is a good measure of storage potential given geologic inputs. Storage area density changes with the plume mass, radius (or area), and $CO_2$ density.

The sensitivity analysis of depth generates generally intuitive results. The increase of $CO_2$ density by increasing pressure has a stronger impact than the decrease in $CO_2$ density by the increasing temperature, leading to an overall increase in $CO_2$ density. Given the same reservoir dimensions, porosity, and permeability, this allows for more $CO_2$ to penetrate the reservoir with increasing depth, leading to higher injection rates, higher reservoir capacity, and fewer wells and space needed to do so. As a result the overall injection and storage cost falls with increasing depth because fewer wells are required, each well can inject more $CO_2$, and the reservoir capacity increases. Similar results have been shown previously using simpler tools such as *CO₂-PENS* [8].

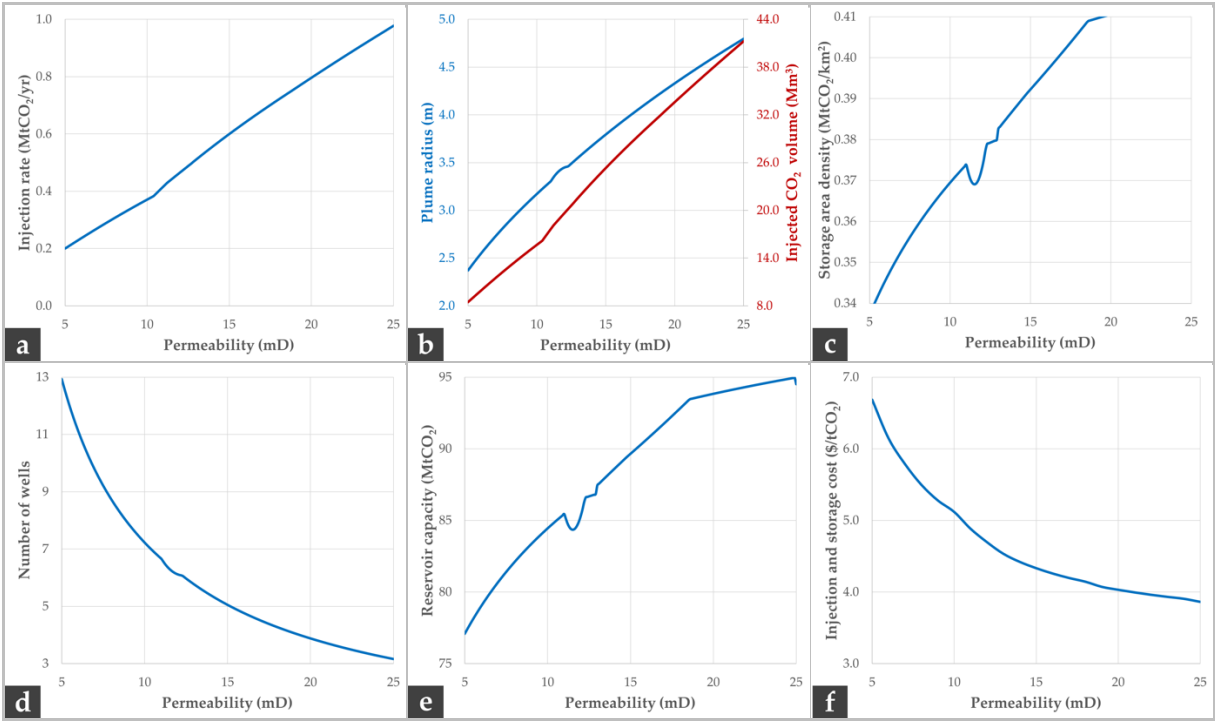

Figure 5: Impact of varying permeability from 5 to 25 mD. Remaining variables: Depth = 2000 m, Thickness = 20 m; porosity = 0.2, geothermal gradient = 25°C/km. By default, *SCO$_2$T* assumes that each reservoir is 256 km$^2$ (roughly 10 by 10 miles). The line kinks in panels (c) and (e) are due to ROM artifacts.

Figure 5 illustrates the impact of holding depth, thickness, porosity, and geothermal gradient constant and varying permeability from 5 to 25 mD. These changes subsequently impact CO$_2$ injection rate (Figure 5a), CO$_2$ plume radius and volume (Figure 5b), storage area density (Figure 5c), number of wells (Figure 5d), reservoir capacity (Figure 5e), and total injection and storage cost (Figure 5f). Changing permeability of the reservoir has a relatively major effect on the injection rate, an intuitive response given permeability has the largest impact on a fluid's ability to flow through a reservoir. When varying permeability, plume radius and injected CO$_2$ volume will largely follow a similar path as injection rate, since increasing permeability will make it more preferential for the plume to expand than to increase pressure in the pore space. It is no surprise that storage area density and reservoir capacity follow similar paths, as lower permeability values make it difficult for the CO$_2$ to penetrate the entirety of the reservoir. The sudden increase in storage area density and reservoir capacity between 10 and 15 mD may be explained by the point at which the CO$_2$ plume is able to effectively reach the majority of the reservoir pore space. Injection and storage costs fall with rising permeability because fewer wells are needed while the reservoir storage capacity increases too. Reservoir storage capacity for *SCO$_2$T* is based on how much CO$_2$ an operator can actually inject/store in a given time period and not a static or arbitrary calculation based available pore space.

Figure 6 illustrates the impact of holding depth, thickness, permeability, and geothermal gradient constant and varying porosity 0.05 to 0.4. These changes can impact CO$_2$ injection rate (Figure 6a), CO$_2$ plume radius and volume (Figure 6b), storage area density (Figure 6c), number of wells (Figure 6d), reservoir capacity (Figure 6e), and total injection and storage cost (Figure 6f). Increasing porosity has no effect on the injection rate (Figure 6a) and therefore has no impact on the injected volume (Figure 6b; secondary y-axis). However, the plume radius does fall as porosity increases since the same volume of $CO_2$ occupies a decreasing proportion of the formation (Figure 6b; primary y-axis). Accordingly, the storage area density also increases because the same amount of $CO_2$ needs a lower 2D footprint to be stored (Figure 6c). Because the plume radius decreases but the injection rate is steady, more wells can be placed in the 2D footprint (Figure 6d) and thus overall reservoir capacity also increases (Figure 6e). Although the number of wells increases across the site—this increases the upfront fixed capital costs—the increased storage area density means that injection and storage costs fall (Figure 6f).

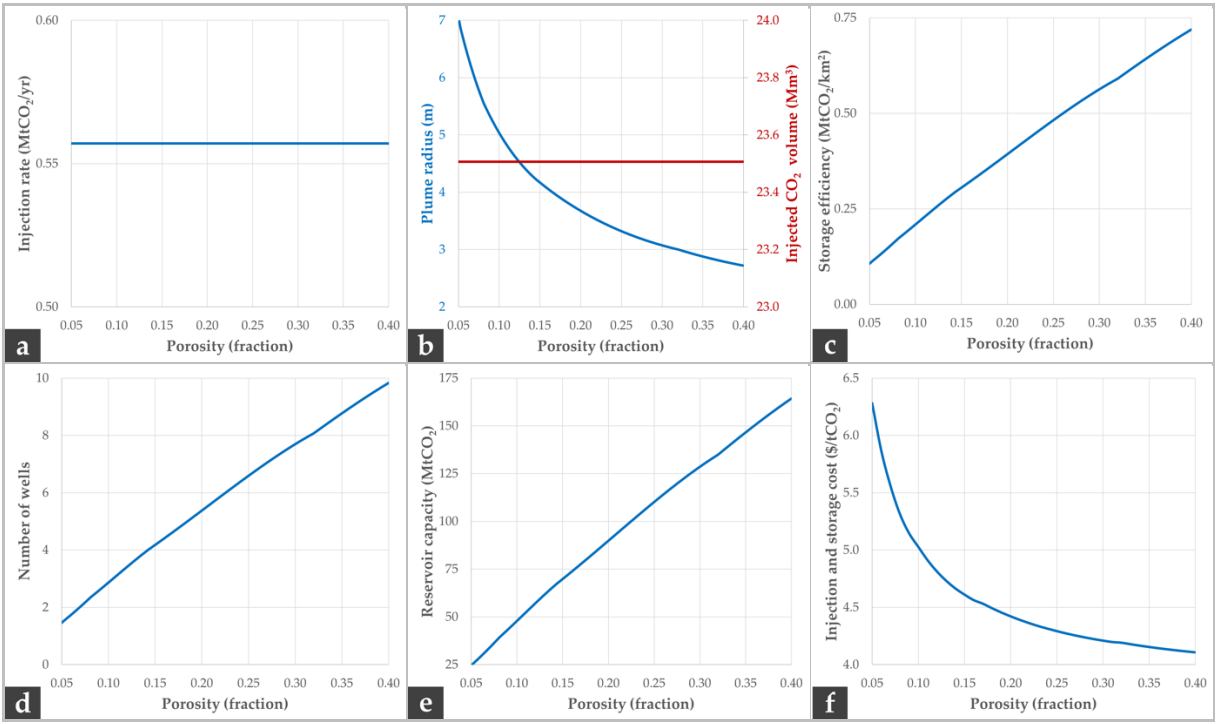

Figure 6: Impact of varying porosity from 0.05 to 0.4 (fraction). Remaining variables: Depth = 2000 m, Thickness = 20 m; permeability = 14 mD; geothermal gradient = 25°C/km. By default, *SCO₂T* assumes that each reservoir is 256 km² (roughly 10 by 10 miles).

## Uncertainty analysis

In addition to understanding the sensitivity of individual geologic parameters across a set range, users can vary multiple parameters at the same time through probabilistic distributions of input parameters. This is shown in Figure 7 where formation thickness, permeability, and porosity are randomly sampled using a normal distribution with a standard deviation equivalent to 10% of the mean value. By default, *SCO₂T* produces scatterplots with up to 10,000 points; users can increase or reduce that limit. In this case, the users can visualize the cloud of possible outcomes on injection rates (top row in Figure 7), plume radius (middle row), and injection and storage costs (bottom row).

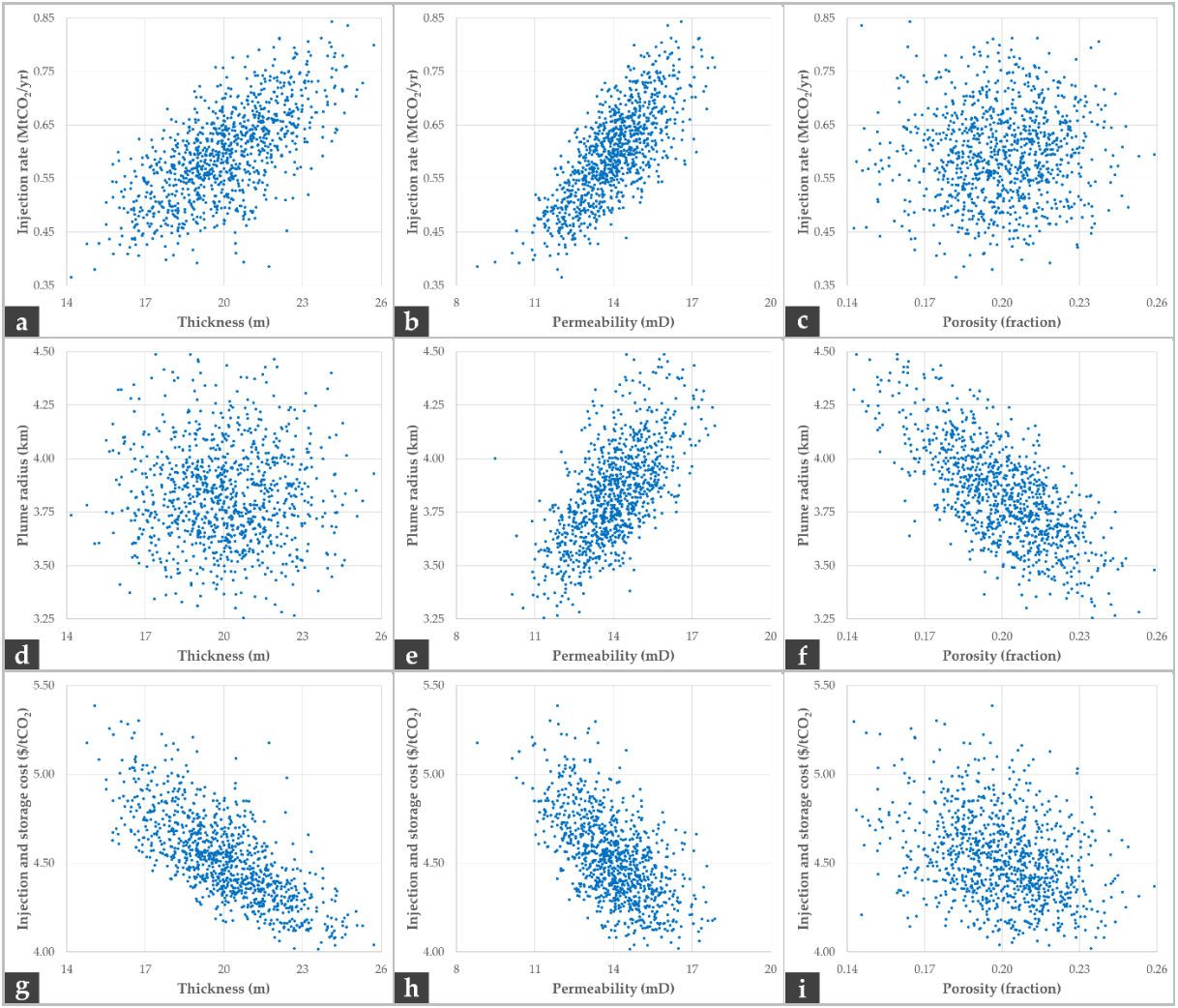

Figure 7: Uncertainty analysis. **Top row:** impact on injection rate from thickness (a), permeability (b), and porosity (c). **Middle row:** impact on plume radius from thickness (d), permeability (e), and porosity (f). **Bottom row:** impact on injection & storage cost from thickness (g), permeability (h), and porosity (i). Uncertainty analysis is based on 1000 simulations based on a normal distribution and standard deviation equivalent to 10% for thickness (mean = 20 m), permeability (mean = 14 mD), and porosity (mean = 0.20). Depth was set to 2000 m and a geothermal gradient of 25°C/km.

Varying several parameters at the same time can lead to results differing from when each parameters is varied independently. For example, increasing thickness increases injection rates (Figure 7a) and reduces costs (Figure 7g) but does not have a meaningful impact on plume radius (Figure 7d) because, in part, permeability (Figure 7e) and porosity (Figure 7f) counteract each other. Permeability, meanwhile, retains a strong forcing over injection rate (Figure 7b), plume radius (Figure 7e) and cost (Figure 7h) even though thickness and porosity are varying. In combination with permeability and thickness, the porosity signal can still be seen for impact on plume radius (Figure 7f) but not for the injection rate (Figure 7c) and final costs (Figure 7i). The total economic impact of randomly varying thickness, permeability, and porosity at the same time can be seen in Figure 8. The histogram highlights the likelihood that the economic performance of a reservoir will fall within a certain range. Unsurprisingly, given that the input variables were assumed to vary with a normal distribution, the output injection and storage costs are largely normally distributed too. *SCO₂T* also allows users to assume a uniform or log-normal distribution for any variables (analysis not shown here).

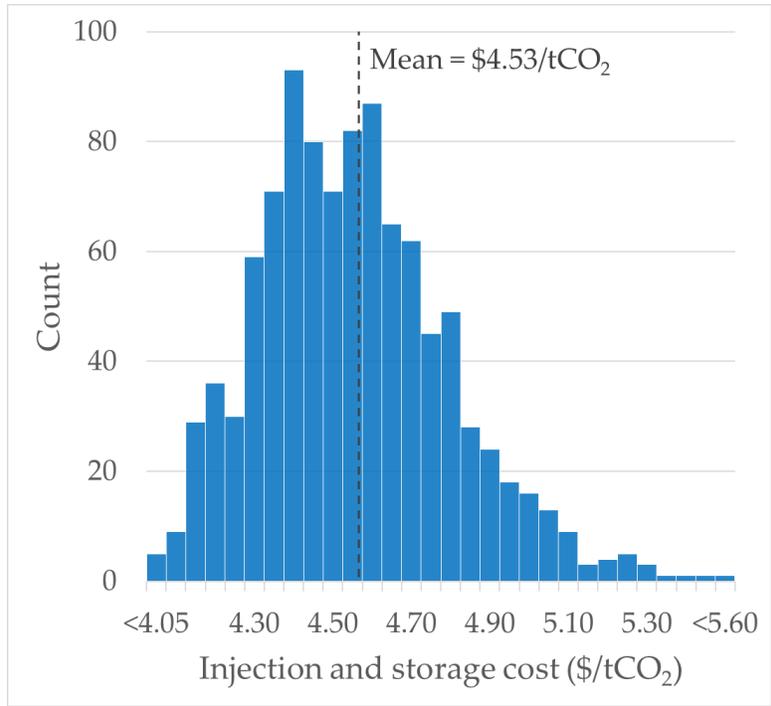

Figure 8: Histogram if $CO_2$ injection and storage costs based on uncertainty analysis of formation thickness, permeability, and porosity.

## CONCLUSIONS AND FUTURE RESEARCH

CCS technology is likely to be widely deployed in coming decades due to both climate and economic drivers. This will require identifying and characterizing suitable storage sites for vast volumes of $CO_2$, including sequestration costs. The *SCO₂T* tool has been developed to address this need including being able to analyze hundreds of thousands or millions of reservoir combinations in only seconds or minutes as well as coupling economics with sequestration engineering.

Through the sensitivity analysis we showed that increasing depth and permeability both can lead to increased $CO_2$ injection rates, increased storage potential, and reduced costs, while increasing porosity reduces costs without impacting the injection rate ($CO_2$ is injected at a constant pressure in all cases) by increasing the reservoir capacity. Through uncertainty analysis—where formation thickness, permeability, and porosity are randomly sampled—we showed that that final sequestration costs are normally distributed with upper bound costs around 50% higher than the lower bound costs. While site selection decisions will ultimately require detailed site characterization and permitting, *SCO₂T* provides an inexpensive screening tool that can help prioritize projects based on the complex interplay of reservoir, infrastructure (e.g., proximity to pipelines), and other (e.g., land use, legal) constraints on the suitability of certain regions for CCS.

Future work will focus on further developing *SCO₂T* and applying it to local-to-national scenarios. Future developments will include adding functionality to explore injection and storage over multiple timeframes as well as over more complex geologies and other parameters including heterogeneity, different fluid properties, different depositional environments, and the ability to fine tune parameters such as residual water and relative endpoint permeabilities. Because *SCO₂T* is so fast running, we also intend to apply *SCO₂T* to national carbon sequestration problems including the spatial coverage of databases including NATCARB [37].

## ACKNOWLEDGMENTS

This research was primarily funded by the U.S. Department of Energy's (DOE) Fossil Energy Office projects "*SimCCS*: Development and Applications" (Award No. FE-1017-18-FY18) and the "Southwest Regional Partnership on Carbon Sequestration" (Award No. DE-FC26-05NT42591). Additional work was funded by DOE's Fossil Energy Office through the US-China Advanced Coal Technology Consortium (Award No. DE-PI0000017) and "Integrated Midcontinent Stacked Carbon Storage Hub" (Award No. FE-0031623), and through the Los Alamos National Laboratory's Laboratory Directed Research Development (LDRD) program.